\documentclass{webofc}
\usepackage[varg]{txfonts}   
\usepackage{graphicx}
\usepackage{xcolor}
\usepackage{subfig}
\usepackage{siunitx}
\usepackage{bm}
\newcommand{\eg}{e.\,g.\@}
\newcommand{\ie}{i.\,e.\@}
\newcommand{\etal}{\emph{et al.}\@}

\newcommand{\Cpg}{\ensuremath{{}^{12}\mathrm{C}(\mathrm{p},\gamma){}^{13}\mathrm{N}}}

\newcommand{\CPG}{\ensuremath{{}^{13}\mathrm{C}(\mathrm{p},\gamma){}^{14}\mathrm{N}}}

\newcommand{\Npa}{\ensuremath{{}^{15}\mathrm{N}(\mathrm{p},\alpha){}^{12}\mathrm{C}}}

\newcommand{\Npg}{\ensuremath{{}^{14}\mathrm{N}(\mathrm{p},\gamma){}^{15}\mathrm{O}}}

\newcommand{\Alpg}{\ensuremath{{}^{27}\mathrm{Al}(\mathrm{p},\gamma){}^{28}\mathrm{Si}}}

\begin{document}
\title{Low-energy Cross Section Measurements of {\boldmath ${}^{\mathsf{12}}\mathsf{C}(\mathsf{p},\gamma)$} Deep Underground at LUNA}
%
%

\author{\firstname{Jakub} \lastname{Skowronski}\inst{1} \and \firstname{Axel} \lastname{Boeltzig} \inst{2} \thanks{\email{a.boeltzig@hzdr.de}} for the LUNA collaboration}

\institute{  Università degli Studi di Padova and INFN, Sezione di Padova, Via Francesco Marzolo 8,  35131 Padova, Italy
        \and Helmholtz-Zentrum Dresden-Rossendorf, Bautzner Landstra\ss{}e 400, 01328 Dresden, Germany}

\abstract{%
  The \Cpg{} reaction cross section is currently under investigation in the low-background environment of the Laboratory for Underground Nuclear Astrophysics (LUNA). It is being studied using different types of solid targets, and employing two complementary detection techniques: HPGe spectroscopy and activation counting. To reduce systematic uncertainties, targets have been accurately characterized and their degradation under the intense beam of the LUNA-400 accelerator monitored. We present the experimental techniques and the corresponding analyses used to extract the reaction cross section.
}
\maketitle
\section{Introduction}
\label{intro}

The \Cpg{} reaction is part of the CNO cycle, which is active in the hydrogen burning regions of main sequence, Red Giant Branch (RGB), and Asymptotic Giant Branch stars (AGB). It does not only contribute to the energy production of the stars, but also governs the abundances of several nuclei in their interiors, \eg{} the $^{12}$C is produced by the \Npa{} reaction and then depleted by proton capture.

The $^{12}$C/$^{13}$C ratio is a useful tracer of nucleosynthesis inside stars~\cite{schoeier2000}, since it can be readily derived by studying the stellar spectra. AGB stars, in particular, are remarkably prolific centers of nucleosynthesis. Unfortunately the AGB phase of stellar evolution is marked by intense mixing phenomena, which make the theoretical models particularly difficult to establish~\cite{herwig2005}. The mixing heavily impacts the internal and external abundances of both the $^{12}$C and $^{13}$C, and can arise from several different sources: convective motion inside the star, the angular momentum of the star~\cite{langer1999}, magnetic buoyancy~\cite{vescovi2020} or gravitational waves~\cite{denissenkov2003}. Hence, by obtaining a precise cross section for the \Cpg{} and the \CPG{} reactions the theoretical models can be improved and will give further insight  into the evolution of AGB stars.

Additionally, the proton capture on $^{12}$C produces $^{13}$N, which is $\beta^{+}$ unstable and one of the main neutrino sources from the CNO cycle. The CNO neutrinos from our Sun have been recently measured for the first time by the Borexino collaboration~\cite{borexino2020}. Since the neutrino rates are directly proportional to the reaction rates, it could be possible to extract the information about the solar metallicity directly from the Borexino result. Up to now, two different estimates of the solar metallicity exists that appear to be in disagreement~\cite{basu2008} creating the so-called Solar Metallicity Problem. Thus a precise value of the \Cpg{} cross section could help in providing a further independent estimate~\cite{gann2021}.

The measurement of the \Cpg{} cross-section at astrophysical energies was the goal of a recently conducted experiment at LUNA. Different experimental methods (targets and detection setups) were employed, in order to characterize and limit systematic experimental uncertainties.

\section{Current Status}
\label{state-of-the-art}

\begin{figure}[htbp]
  \centering
  \includegraphics[width=14cm,clip]{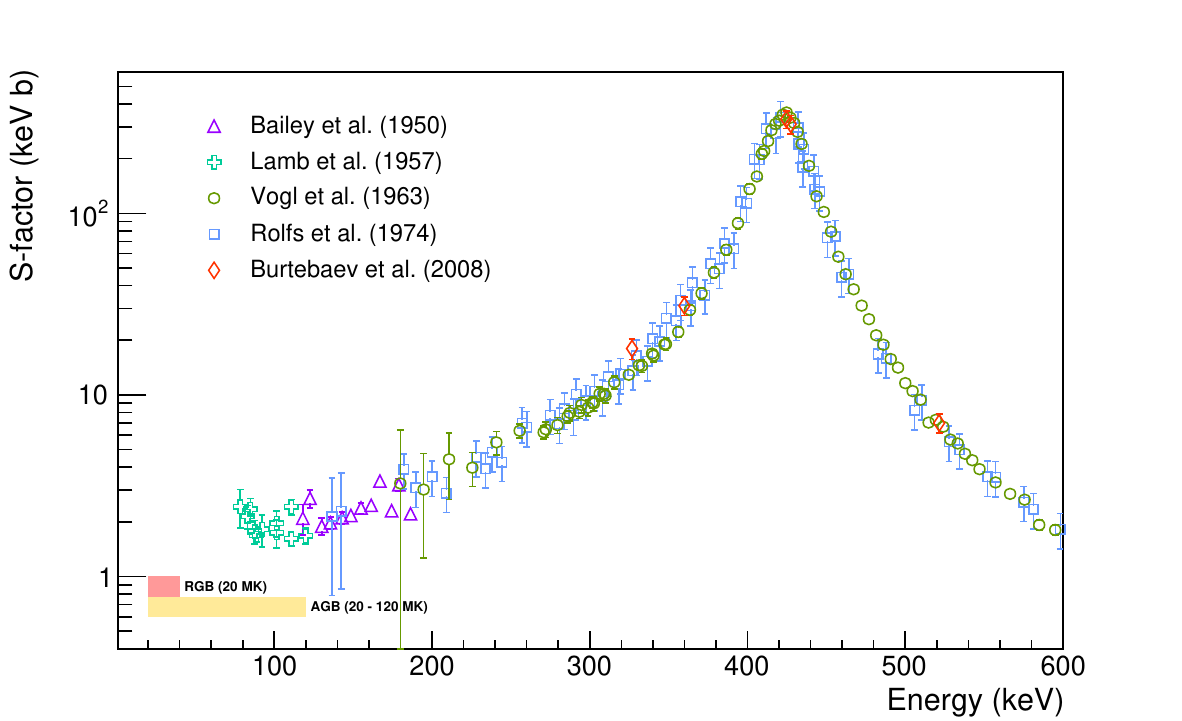}
  \caption{State of the art for the \Cpg{} reaction cross section. Two different energy ranges of interest are highlighted for RGB and AGB stars, respectively. Direct data available in these energy ranges comes from early measurements with significant scatter and uncertainties.}
  \label{fig-lit}
\end{figure}

Several previous studies of the \Cpg{} reaction cross section are available, with their results summarized in Fig.~\ref{fig-lit}. There are two complete studies that report the cross-sections down to \qty{200}{\kilo\electronvolt}, performed by Vogl~\cite{vogl1963}, and Rolfs \etal{}~\cite{rolfs1974}. Both data sets covered a broad energy region and are generally in good agreement, but show a discrepancy in the precise value of the energy of the observed broad resonance. Detection of prompt $\gamma$-rays was used in both these experiments, using thin evaporated targets. A recent measurement by Burtebaev \etal{} \cite{burtebaev2008} used a thick graphite disk, and obtained data points at relatively high energies, found to be in agreement with the previous data sets. In the region of astrophysical interest, however, only older measurements are available: Bailey and Stratton \cite{bailey1950} detected charged particles from the decay of ${}^{13}$N, whereas Lamb and Hester \cite{lamb1957} measured prompt $\gamma$-rays from the reaction. Both of these measurements used thick graphite targets. Their results for the cross section at low energies have large uncertainties, and show significant scatter. 

\section{Experimental Setup}
\label{exp-setup}

The LUNA-400\,kV accelerator was chosen for this measurement, as it provides proton beams with high intensity (up to $\simeq \qty{400}{\micro\ampere}$ in this experiment) at excellent beam energy resolution and stability \cite{formicola2003}. Its deep-underground location at the Gran Sasso National Laboratory (LNGS) results in a drastic reduction of cosmic backgrounds, thus improving experimental sensitivity.

Two types of targets were used for the study of \Cpg{}: thin targets produced by evaporation of carbon powder onto Tantalum backings, and thick targets of solid graphite. The graphite targets were sufficiently thick to stop the proton beam, providing the conditions for the measurement of the infinitely-thick-target yield. The carbon in both targets had the natural isotopic composition, \ie{}, containing about \qty{1}{\percent} of ${}^{13}$C. Targets were characterized on-site by analyzing the width and shape of the primary $\gamma$-ray peaks measured with the HPGe detector \cite{ciani2020}. Target stability was further monitored in-situ by repeated measurements at a given reference proton energy. After the measurements at LUNA, targets were characterized by NRRA at the Institute for Nuclear Research (ATOMKI).

\begin{figure}[htbp]
  \centering
  {\includegraphics[height = 38mm]{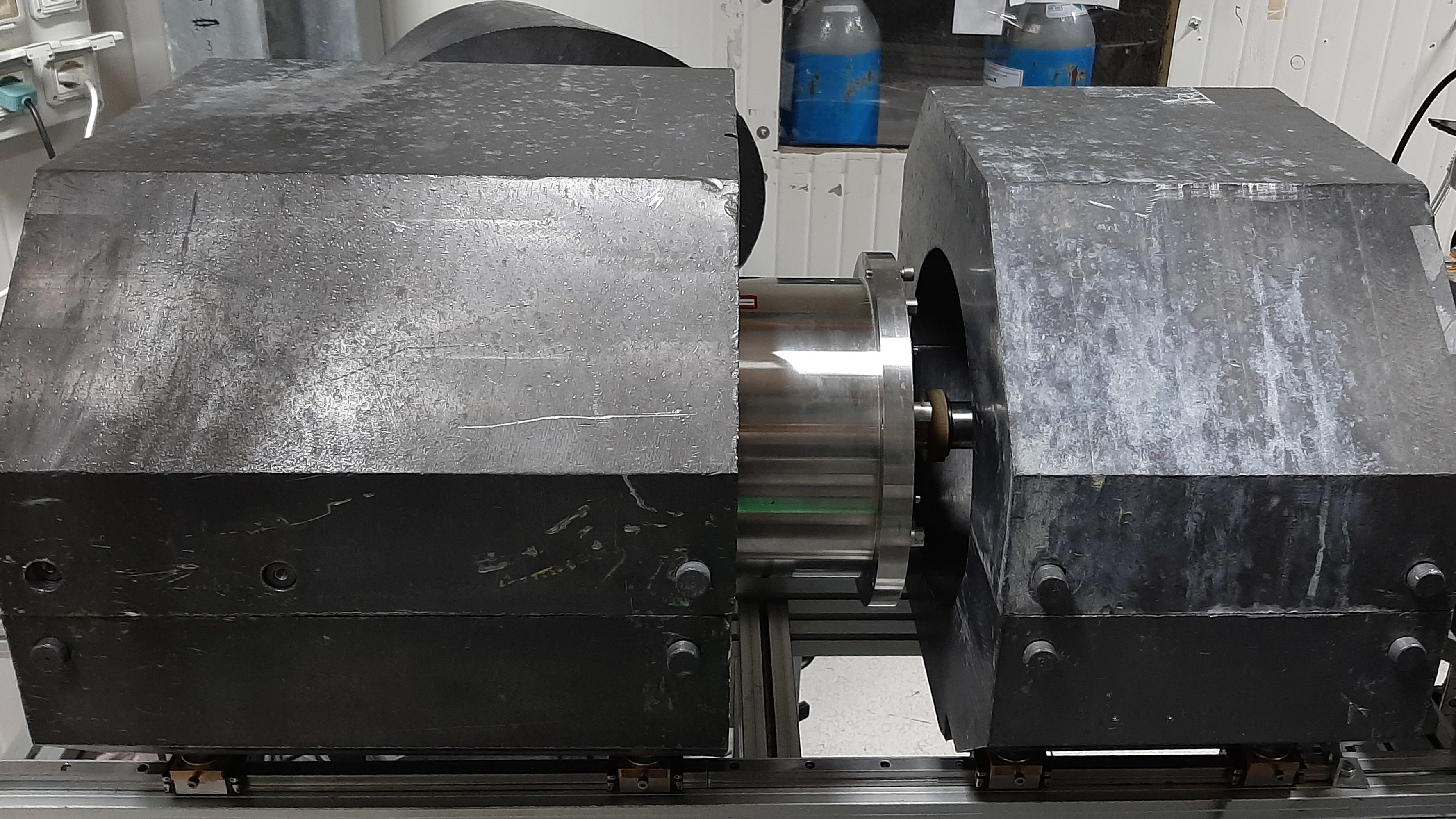}}
  \hspace{3em}
  {\includegraphics[height = 38mm]{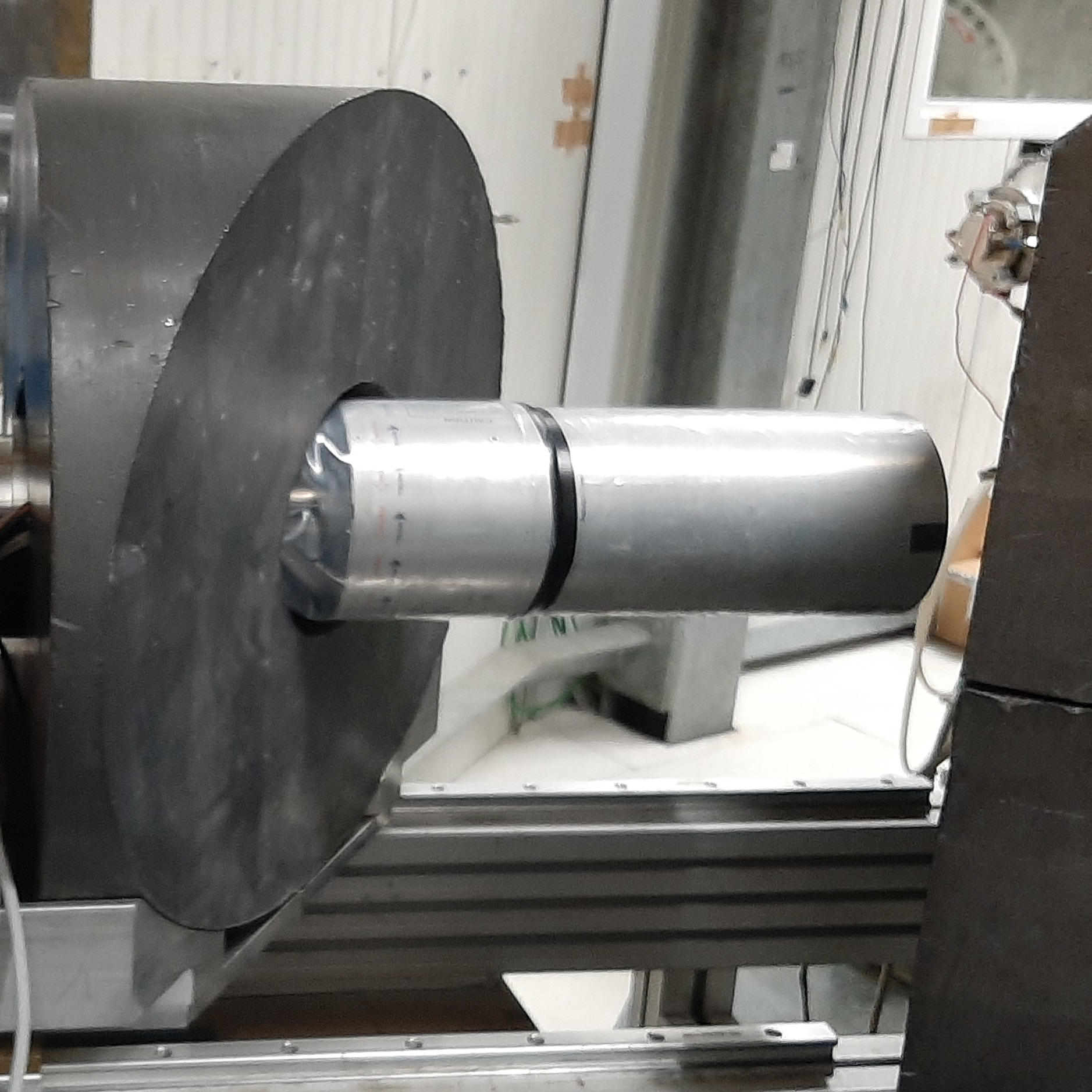}}
  \vspace*{-0.1cm}
  \caption{Left: BGO detector in lead shielding, partially retracted (shielding open) to access target. Right: HPGe detector, here shown retracted (fully outside of the shielding) at \qty{55}{\degree} to the target.  }
  \label{fig:setup}
  \vspace*{-0.3cm}
\end{figure}

For the detection of the \Cpg{} reaction, two detectors were used in separate experimental campaigns (Fig.~\ref{fig:setup}): a High-Purity Germanium detector (HPGe), and a Bismuth Germanium Oxide (BGO) detector. The HPGe detector was placed at \qty{0}{\degree} or \qty{55}{\degree} towards the target. With its excellent energy resolution it allowed for $\gamma$-ray spectroscopy of the single primary $\gamma$-ray of the reaction. With the reaction $Q$-value of \qty{1943.5(3)}{\kilo\electronvolt}, the primary $\gamma$-ray energy region is affected by environmental $\gamma$-rays in the measured spectrum. A \qty{15}{cm} lead shielding \cite{boeltzig2018} around target and detector setup was used to reduced these backgrounds. The large segmented BGO detector \cite{casella2002}, providing almost $4\pi$ coverage around the target, was used to detect the reaction in two ways: i) by detecting the primary $\gamma$-ray, and ii) by detecting the signature of annihilation $\gamma$-rays from the $\beta^+$-decay of ${}^{13}$N. The former method provides a large detection efficiency and is rather insensitive to angular distribution effects, thanks to the large solid angle covered by the detector. For this reaction, however, it is limited in sensitivity, owing to the non-negligible background of the BGO detector in the energy region of interest (largely given by intrinsic radioactivity \cite{boeltzig2018}), and could only be used at the higher energies of this experiment. For the second method, the signature of two $\gamma$-rays with an energy of \qty{511}{\kilo\electronvolt} in opposite segments of the detector was exploited to detect these low-energy events in the detector background. Activation (beam-on) and counting (beam-off) phases were run in the same experimental configuration, \ie{}, counting was performed in-situ.

\section{Data Analysis}
\label{ana-methods}

\begin{figure}[htbp]
  \centering
  \vspace{-0.3cm}
  \includegraphics[width=13cm,clip]{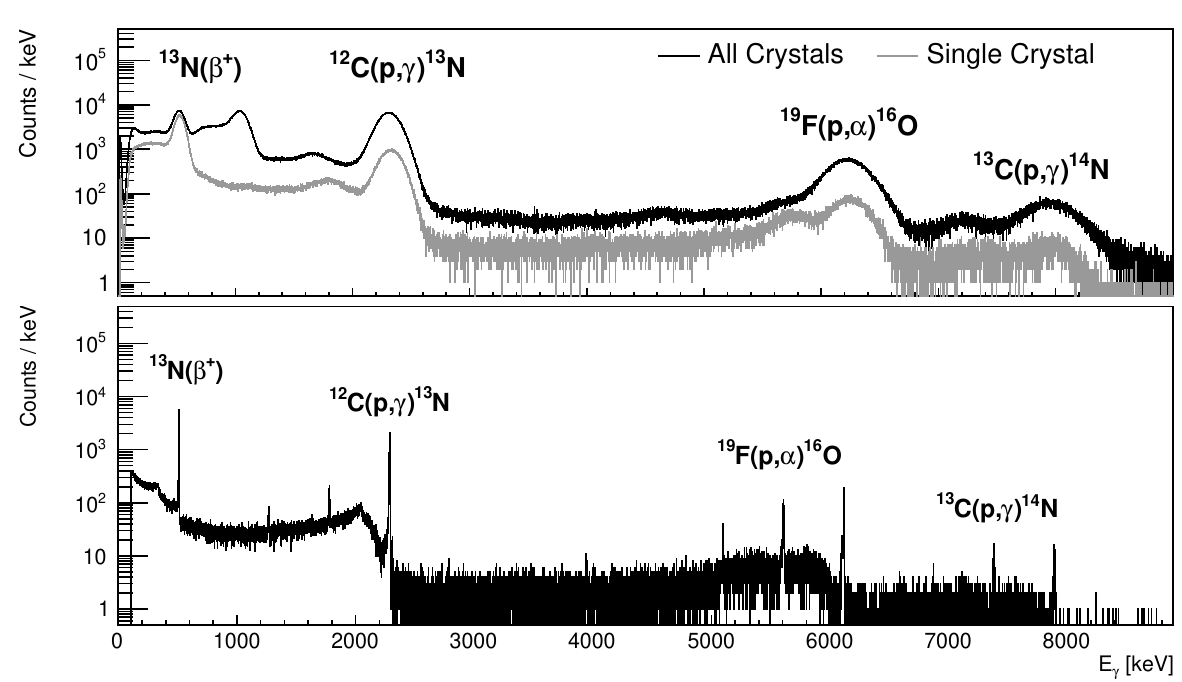}
  \caption{Examples of the spectra acquired at $E_{p} = 380$~keV. The upper panel shows the spectrum from the BGO. The lower panel shows the HPGe spectrum.}
  \label{fig-spectra}
\end{figure}

The data taken were analyzed with techniques corresponding to the characteristics of the two detection setups. Examples for spectra taken with both detectors are shown in Fig.~\ref{fig-spectra}. In case of the HPGe campaign, the primary $\gamma$-peak was parameterized as described in~\cite{ciani2020}. By fitting to the experimental data, the target degradation could be studied and the $S$-factor could be directly extracted from the $\gamma$-peak fitting procedure. Additionally, the target degradation obtained from the fits was compared with the results of the NRRA analysis. This approach was only followed for the case of the thin target. In case of the thick graphite disk, the yield was extracted from the primary $\gamma$-peak and compared to the results for the thin targets. The efficiency was obtained with a multiparametric fit to the data obtained with $^{137}$Cs, $^{60}$Co and $^{133}$Ba calibration sources and resonances in \Npg{} \cite{imbriani_s-factor_2005,daigle_measurement_2016} and \Alpg{} \cite{iliadis_low-energy_1990} with well-known strengths and branching ratios. The majority of the HPGe data was taken at \qty{0}{\degree}, whilst measurements at \qty{55}{\degree}, allowed to check and confirm the expected angular distribution of the primary $\gamma$-ray.

For the BGO data, the major part of the analysis focused on the $\beta^{+}$ decay of $^{13}$N from the reaction. Since the detector is segmented in 6 different crystals, it was possible to exploit the distinct signature of the \qty{511}{keV} $\gamma$-rays from positron annihilation by searching for two $\gamma$-rays in opposite crystals, thus reducing the background significantly. Since the target was inside the detector both during the irradiation and the counting, the full activation curve was analyzed by fitting the data to the solution of the differential equation that governs the amount of $^{13}$N inside the target. This approach not only permitted the measurement of the cross-section down to \qty{70}{keV} (center-of-mass) but also to check for any potential contaminant in the target. In this case, only thick graphite disks were employed in order to maximize the $^{13}$N production. Additionally, the prompt-$\gamma$ peak was analyzed for runs with $E_\mathrm{p} > \qty{180}{keV}$, where the signal was still above the background. At these energies, the yields from prompt-$\gamma$ and activation measurements were compared and found to be compatible. The efficiency for the activation measurements was determined by using the \Npg{} reaction, measuring the $\beta^+$ decay of $^{15}$O. Additionally, the activation efficiency was verified by ex-situ counting of an activated graphite sample with an HPGe detector. The obtained efficiency value was further validated with the use of a simulation based on Geant4. In the simulation, the positron energy spectrum of the both decays studied here, ${}^{13}$N and ${}^{15}$O, was included; the differences in efficiency between the positrons of both nuclei were found to be small compared to other factors such as a variation of the position of the beam spot on target.

The data analysis for the measurement of \Cpg{} at LUNA-400 has been completed. The results of the different parts of this experiment are found to be in good agreement between the different targets and detection setups. The obtained cross section data will be presented in a forthcoming publication.





\section*{Acknowledgements}

D.\ Ciccotti, the technical staff and mechanical workshops of the LNGS and INFN Divisions of Padua and Naples are gratefully acknowledged for their support for this experiment.
This work was supported by INFN. 
J.\,S.\ acknowledges support by the Italian Ministry of Education, University and Research (MIUR) through the ``Dipartimenti di eccellenza'' project ``Physics of the Universe''.
%
%
%
%
%
A.\,B.\ was supported by
the European Union (
ERC-StG \emph{SHADES}, no. 852016; and \emph{ChETEC-INFRA}, no. 101008324). 

\bibliography{12C_pg_NPA.bib}

\end{document}